\documentclass[aps,showpacs,showkeys,preprint]{revtex4}
\usepackage[dvips]{graphicx}
\usepackage{latexsym,amsmath,amssymb}

\begin{document}

 \title{Regular black hole in three dimensions}

\author{Yun Soo Myung}
  \email{ysmyung@inje.ac.kr}
  \affiliation{Institute of Basic Science and School of
    Computer Aided Science, \\ Inje University, Gimhae 621-749, Korea}
\date{\today}

\author{Myungseok Yoon}
  \email{younms@sogang.ac.kr}
  \affiliation{Center for Quantum Spacetime, Sogang University, Seoul
    121-742, Korea}
\date{\today}

\begin{abstract}

We find a new  black hole in three dimensional anti-de Sitter
space by introducing an anisotropic perfect fluid inspired by the
noncommutative  black hole. This is a regular black hole with two
horizons. We compare thermodynamics of this black hole with that
of non-rotating BTZ black hole.  The first-law of thermodynamics
is not compatible with the Bekenstein-Hawking entropy.

\end{abstract}


\maketitle


\section{Introduction}
\label{sec:intro} Hawking's semiclassical analysis of  the black
hole radiation suggests that most  information about initial
states is shielded behind the event horizon and will not back to
the asymptotic region far from the evaporating black
hole~\cite{HAW1}. This means that the unitarity is violated by an
evaporating black hole. However, this conclusion has been debated
by many authors for three decades~\cite{THOO,SUS,PAG}. It is
closely related  to a long standing puzzle of  the information
loss paradox, which states the question of whether the formation
and subsequent evaporation
 of a black hole is unitary. One of the most urgent problems in black
hole physics is  to resolve the unitarity issue. In this
direction, a complete description of black hole evaporation is an
important issue. In order to determine the final state of
evaporation process, a more precise treatment including
 quantum gravity effects  and backreaction is generally required. At present, two
leading candidates for quantum gravity are the string theory and
the loop quantum gravity. Interestingly, the semiclassical
analysis of the loop quantum black hole provides a regular black
hole (RBH) without singularity whose minimum size $r_c$ is at
Planck scale $l_{p}$,  in contrast to the classical
one~\cite{MOD}.

RBHs have been considered, dating back to Bardeen~\cite{BAR}, for
avoiding the curvature singularity beyond the event horizon in
black hole physics~\cite{RBH}. Their causal structures are similar
to the Reissner-Nordstr\"{o}m  black hole with the singularity
replaced by  de Sitter space-time with curvature radius
$\tilde{r}_0=\sqrt{3/\Lambda}$~\cite{Dymn,Ans}.  Hayward has
discussed the formation and evaporation process of a RBH with
minimum size $l$~\cite{HAY} which can be identified with the
minimal length induced  from the string theory~\cite{Vene}, and
its thermodynamic analysis was performed in~\cite{MKYr}. A
rigorous treatment of the evaporation process was carried out for
the renormalization group (RG) improved black hole with minimum
size $r_{cr}=\sqrt{\tilde{\omega}G}$~\cite{REU}.

On the other hand,  the noncommutativity  with parameter $\theta$
may provide another RBH with minimum scale $\sqrt{\theta}$:
noncommutative black hole~\cite{SS,nicol,nicol2} and its
commutative limit is the Schwarzschild black hole. Recently, the
authors~\cite{MKP} have investigated thermodynamics and
evaporation process of this noncommutative black hole. The
thermodynamics similarity between the noncommutative and
Reissner-Nordstr\"{o}m black holes was shown in Ref.\cite{KSY}.
 The entropy
issue of this black hole was discussed in~\cite{BMS,BMM} and the
Hawking radiation was considered in~\cite{NM}. The connection
between their minimum sizes is given by $r_c \sim \tilde{r}_{0}
\sim l \sim r_{cr} \sim \sqrt{\theta}\sim l_{p}$.

In this work, we construct a new  black hole in AdS$_3$ spacetimes
by introducing an anisotropic perfect fluid inspired by the 4D
noncommutative  black hole. This is a regular black hole with two
horizons in three dimensions. We compare thermodynamics of this
black hole with that of non-rotating BTZ black hole  (NBTZ). The
first-law of thermodynamics is not compatible with the
Bekenstein-Hawking entropy. Finally, we discuss thermodynamics of
3D noncommutative black holes based on the Gaussian distribution.

\section{3D regular  black hole }
\label{sec:RBH} We start with a cylindrically symmetric line
element in three dimensions
\begin{equation}
  ds^2 = -f(r) dt^2 + f(r)^{-1} dr^2 + r^2 d\phi^2,
\end{equation}
where $f$ is the metric function to be determined.

It has been shown~\cite{SS} that the noncommutativity eliminates
point-like structures in favor of smeared objects in flat
spacetime. A way of implementing the effect of smearing is a
substitution rule: in four-dimensional (4D) spacetimes,
Dirac-delta function $\delta_{4D}(r)$ is replaced by a Gaussian
distribution of the minimal width
$\sqrt{\theta}$~\cite{SS,MKP,nicol} as
\begin{equation} \label{4Dgd}
\rho^{4D}_{\theta}(r)=\frac{M}{(4\pi \theta)^{3/2}}
e^{-r^2/4\theta}
\end{equation}
whose mass distribution is defined by
\begin{equation} \label{4Dmd}
m^{4D}_{\theta}(r)=\int^r_04 \pi r'^2 \rho^{4D}_{\theta}(r)dr'=
\frac{2M}{\sqrt{\pi}}\gamma(3/2,r^2/4\theta).
\end{equation}
Here $\gamma(3/2,r^2/4\theta)$ is the lower incomplete gamma
function defined as defined by
\begin{equation}
  \gamma(a,z) = \int_0^z t^{a-1} e^{-t} dt.
\end{equation}
In the limit of $r^2/4\theta \to \infty$, one finds
$m^{4D}_{\theta}\to M$.

In three dimensions, Dirac-delta function $\delta_{3D}(r)$ is
replaced by a Gaussian distribution of the minimal width
$\sqrt{\theta}$ as
\begin{equation} \label{3Dgd}
\rho^{3D}_{\theta}(r)=\frac{M}{4\pi \theta} e^{-r^2/4\theta}
\end{equation}
whose mass distribution is simply calculated to be
\begin{equation} \label{3Dmd}
m^{3D}_{\theta}(r)=\int^r_02 \pi r' \rho^{3D}_{\theta}(r)dr'=
M\gamma(1,r^2/4\theta)=M(1-e^{-r^2/4\theta}).
\end{equation}
In the limit of $r^2/4\theta \to \infty$, one recovers
$m^{3D}_{\theta}\to M$ easily.

The Gaussian distribution (\ref{3Dgd})  may be suitable for
describing a three-dimensional (3D) noncommutative black hole.
However,  we will show in Sec. VI that this choice makes an
difficulty to define a small black hole. It turns out that the 3D
noncommutative black hole does not have two horizons and it takes
a degenerate horizon  at the origin $r=0$.  However, the small 3D
noncommutative black hole is not defined in the limit of $r_H\to
0$ because  a smeared (Gaussian)  distribution around the origin
is not appropriate to make a small black hole.

On the other hand, the 4D noncommutative black hole has two
horizons and thus it becomes an extremal black hole for $r_C=r_E$.
As far as concerned on the study of thermodynamics of black holes,
the relevant region to observer at infinity is outside the
degenerate horizon. Hence it is promising to obtain  a 3D black
hole with two horizons. For this purpose, we wish to look for a
different mass distribution which may offer $m_\theta^{4D}$ in
Eq.(\ref{4Dmd}).

 To this end, we compare  the 4D Poisson equation of
$\partial^2_r (1/r) \sim \delta_{4D}(r)$ with 3D equation
$\partial^2_r (\ln r) \sim \delta_{3D}(r)$. Considering the
relation of $\partial_r \ln r=1/r$, a quantity of $\partial_r
e^{-r^2/4\theta} \sim 2 re^{-r^2/4\theta}$ is a similar object in
3D spacetimes. Hence,
 we introduce a new mass density of  cylindrically symmetric,
smeared gravitational source as
\begin{equation}  \label{3Dgdn}
  \rho_\theta(r) = \frac{Mr}{4(\pi\theta)^{3/2}} \exp\left(-
    \frac{r^2}{4\theta}\right)
\end{equation}
whose mass distribution mimics the 4D mass distribution
$m^{4D}_{\theta}$ as
\begin{equation} \label{3Dmdn}
m_{\theta}(r)=\int^r_02 \pi r'
\rho_{\theta}(r)dr'=\frac{2M}{\sqrt{\pi}}\gamma(3/2,r^2/4\theta).
\end{equation}
We note that this mass distribution  differs from the
$m^{3D}_{\theta}$ in Eq.(\ref{3Dmd})  for the 3D noncommutative
black hole.   This choice is meaningful for a 3D smeared
gravitational source because it provides a regular black hole with
two horizons. For another 3D noncommutative BTZ black hole, see
Ref.\cite{KPRY}.

\begin{figure}[pt]
  \includegraphics[width=0.5\textwidth]{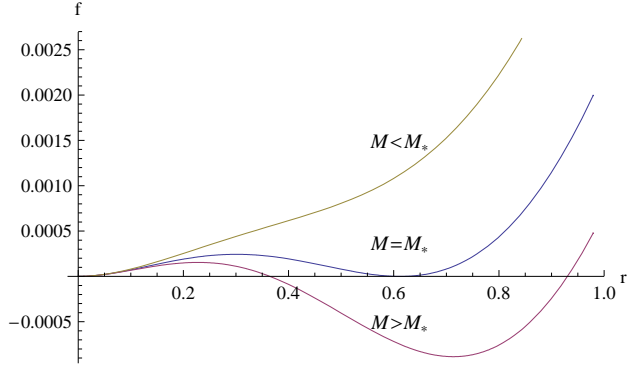}
  \caption{Metric function $f$ as function of $r$ with $\theta = 0.1$
    and $\ell = 10$.  For $M=M_* = 0.0012$, the degenerate horizon is
    located at $r_* = 0.6121$, while for $ M>M_*$, the black hole
    appears with the inner horizon $r_C = 0.3647$ and the outer horizon
    $r_H = 0.9304$. For $M<M_*$, there is no black hole.}
  \label{fig:f}
\end{figure}

In order to find a black hole solution in AdS$_3$ spacetime, we
introduce the Einstein equation
\begin{equation}
R_{\mu\nu}-\frac{1}{2}R g_{\mu\nu} = 8\pi T_{\mu\nu} + \Lambda
g_{\mu\nu},~{\rm with}~ \Lambda = 1/\ell^2,
\end{equation}
where the energy-momentum tensor takes an  anisotropic form
\begin{equation} \label{emt}
 T^\mu~_\nu = {\rm diag}( -\rho_\theta, p_r, p_\perp).
\end{equation}
The Bianchi identity is satisfied ($T^\mu~_\nu$ is conserved)  if
the radial pressure $p_r$ and tangential pressure $p_\perp$
satisfy the relations, respectively
\begin{equation}
p_r = -\rho_\theta,~ p_\perp = -\rho_\theta - r \rho_\theta~',
\end{equation}
where $'$ denotes the derivative with respect to $r$.
 Components of Einstein equation are given by
\begin{eqnarray}
  (tt)~{\rm or}~(rr): \frac{f'}{2r} &=& -8\pi \rho_\theta  + \frac{1}{\ell^2} \\
  (\phi\phi): \frac12 f'' &=& 8\pi  p_\perp + \frac{1}{\ell^2}.
\end{eqnarray}
Solving the above equations determines the metric function to be
\begin{equation}
  f(r,\theta) = \frac{r^2}{\ell^2} - \frac{16M}{\sqrt{\pi}} \gamma\left(
    \frac32, \frac{r^2}{4\theta} \right)=\frac{r^2}{\ell^2}-8m_{\theta}(r).
\end{equation}
We note that $r^2/4\theta \to \infty$, when either $r\to \infty$
or $\theta \to 0$. The former corresponds to the large black hole,
while the latter corresponds to the commutative limit.
 In the limit of $r^2/4\theta \to \infty$, one finds
$\gamma(3/2,\infty)=\sqrt{\pi}/2$  which leads to the metric
function for the NBTZ
\begin{equation}
f_{NBTZ}(r)=\frac{r^2}{\ell^2} - 8M.
\end{equation}
In order to obtain a black hole solution, we find the solution to
$f=0$ numerically. Two horizons come together at the value of mass
$M=M_*$, which puts a lower limit on the black hole mass.  For
 given $\theta$, the horizon of extremal black hole is determined from the
conditions of $f=0$ and $f'=0$ as
\begin{equation}
\frac{r_*}{2\sqrt{\theta}} = \alpha, \end{equation} where $\alpha=
0.9679 $ is the value satisfying
\begin{equation}
  \label{alpha}
  \gamma\left(\frac32, \alpha^2 \right) = \alpha^3 e^{-\alpha^2}.
\end{equation}
This implies that the minimal length of regular black hole is $r_*
\simeq 2 \sqrt{\theta}$.  As is shown in Fig.~\ref{fig:f}, for
$M<M_*$ there is no solution to $f=0$ while for $M>M_*$ there
exist two horizons: the cosmological horizon $r_C$ and event
horizon $r_H$.
\begin{figure}[pt]
  \includegraphics[width=0.5\textwidth]{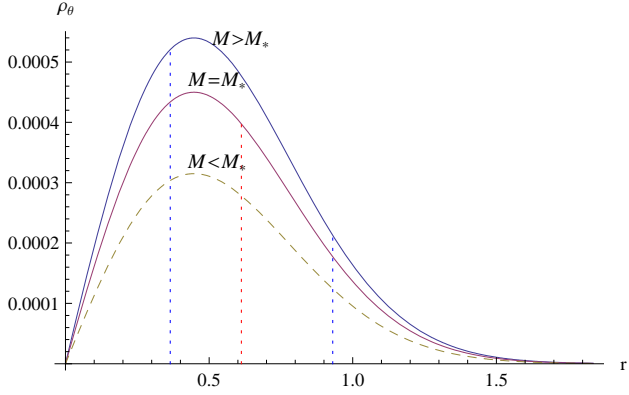}
  \caption{The density profiles $\rho_\theta(r)$ are plotted for
    different mass $M$. When
    $M<M_*$, the black hole cannot be formed (dashed line).
    There is an extremal black hole with horizon at $r=r_*$ for
    $M=M_*$ and a black hole with the inner horizon ($r_C=0.3647$) and outer
    horizon ($r_H=0.9304$) for the mass $M=0.0014$.}
  \label{fig:rho}
\end{figure}
The corresponding density profiles are depicted in
Fig.~\ref{fig:rho}. Then, the mass of the extremal black hole is
determined from the condition of $f(r_*,\theta)=0$ to be
\begin{equation}
  M_*(\theta) = \frac{\sqrt{\pi}  e^{\alpha^2} }{4\alpha
    \ell^2}\theta.
\end{equation}
\begin{figure}[pt]
  \includegraphics[width=0.5\textwidth]{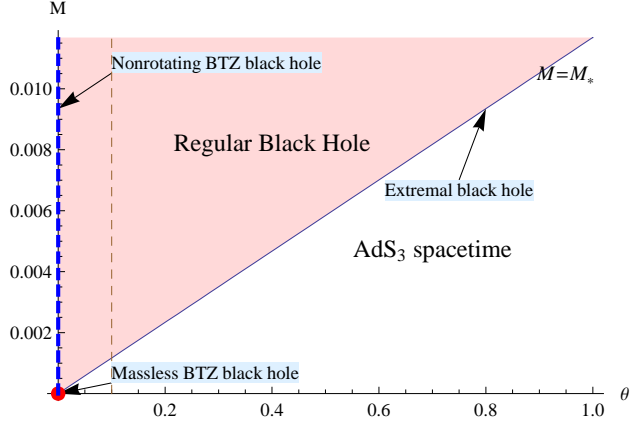}
  \caption{Mass $M$ versus $\theta$ with $\ell = 10$.  The extremal
    black hole is described by $M=M_*(\theta)$. For given $\theta$,
    the regular black holes appear for $M>M_*$, while for $M<M_*$, the
    spacetimes is just the pure AdS$_3$ space.  The $M$-axis
    ($\theta=0$) represents the NBTZ.}
  \label{fig:Mtheta}
\end{figure}
The whole picture is given by Fig.~\ref{fig:Mtheta}. At
$\theta=0$, we find a massless BTZ black hole with $M=0$ and the
NBTZ with $M \not=0$. In case of $\theta \not=0$ (dashed vertical
line), we have regular black hole for $M>M_*$, extremal black hole
at $M=M_*$ and AdS$_3$ spacetime for $M<M_*$.

\begin{figure}[pt]
  \includegraphics[width=0.5\textwidth]{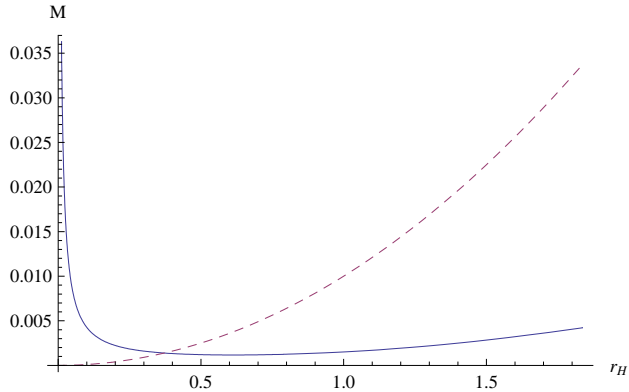}
  \caption{Mass versus $r_C$ and $r_H$. Solid curve: for $r\le
    r_*=0.6121$, one uses $r=r_C$, while for $r\ge r_*$, one uses
    $r=r_H$. The minimum mass $M=M_* $ occurs at
    $r_C=r_H=r_*$. Dashed curve represents the mass as function of
    $r_H$ for the NBTZ.}
  \label{fig:M}
\end{figure}

From the condition of $f=0$, the mass function of horizon radii
$r_C$ and $r_H$ is given by
\begin{equation}
  M(r_{C/H},\theta) = \frac{\sqrt{\pi}}{16 \ell^2}\frac{r_{C/H}^2}{ \gamma(\frac{3}{2},\frac{r_{C/H}^2}{4\theta})}.
\end{equation}
In the limit of $\theta \to 0$, one finds  the mass function for
NBTZ as
\begin{equation}
  M(r_H,\theta \to 0) = \frac{r_H^2}{8\ell^2}.
\end{equation}
These are depicted in Fig.~\ref{fig:M}.

Three conditions for the existence of regular black hole in
AdS$_3$ spacetimes are checked~\cite{Dymn}: i) regularity of the
metric function $f(r)$ and  energy density $\rho_{\theta}(r)$ at
the origin of coordinate $r=0$.  ii) asymptotically AdS spacetimes
and the finiteness of ADM mass ($M<\infty$). iii) dominated energy
condition for the energy-momentum tensor $T_{\mu\nu}$ in Eq.
(\ref{emt}). However, from Fig.~\ref{fig:rho}, one finds that
$\rho_\theta~'>0$ for $r<r_m$ and $\rho_\theta~'<0$ for $r>r_m$,
where $r_m=\sqrt{2\theta}<r_*$ is the maximum value determined by
$\rho_\theta~'=0$. The dominant energy condition of
$T^{00}\ge|T^{ab}|(a,b=1,2)$ is equivalent to $\rho_\theta\ge
0\wedge-\rho_\theta \le p_r \le \rho_\theta \wedge -\rho_\theta
\le p_{\perp} \le \rho_\theta$~\cite{Ans}. Hence, $T_{\mu\nu}$
violates the dominant energy condition for $r<r_m$. The weak
energy condition of $T_{\mu\nu}\xi^\mu\xi^\nu\ge 0$ for any
timelike vector $\xi^\mu$ is equivalent to $\rho_\theta\ge
0\wedge\rho_\theta+ p_r \ge 0 \wedge \rho_\theta+ p_{\perp} \ge
0$. Also, $T_{\mu\nu}$ violates the weak energy condition for
$r<r_m$. Finally, the strong energy condition of
$\rho_\theta+p_r+p_{\perp}\ge 0\wedge\rho_\theta+ p_r \ge 0 \wedge
\rho_\theta+ p_{\perp} \ge 0$ is violated for $r<r_m$.

\section{Thermodynamics of 3D regular black hole}
From the condition  of  $T_H=f'(r_H,\theta)/4\pi$, we obtain the
Hawking temperature
\begin{equation}
  T_H(r_H,\theta) = \frac{r_H}{2\pi\ell^2} \left[ 1 - \left(\frac{r_H^2}{4\theta}
    \right)^{\frac32} \frac{\exp \left( - \frac{r_H^2}{4\theta}
      \right)}{\gamma\left(\frac32, \frac{r_H^2}{4\theta}
        \right)} \right]. \label{T}
\end{equation}
For $\theta=0.1$, we have the temperature, showing the deviation
from $T_H(r_H,\theta \to 0)=r_H/2\pi \ell^2$ of the NBTZ  for
small $r_H$. The temperature is a monotonically increasing
function of horizon radius for large $r_H$. Also, we
    observe that $T_{H}(r_*,\theta)=0$ at $r_H=r_*$, indicating the
    zero temperature for the extremal black hole.
\begin{figure}[pt]
  \includegraphics[width=0.5\textwidth]{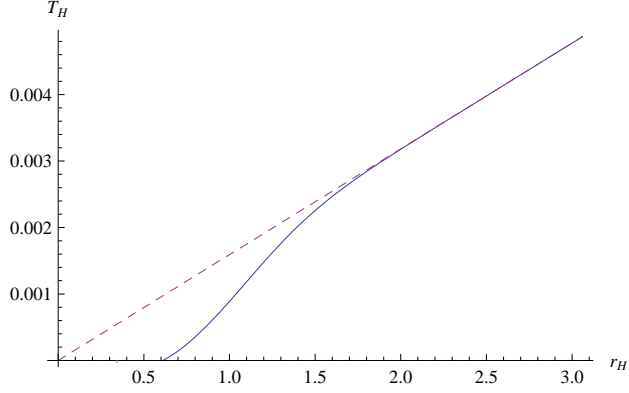}
  \caption{Hawking temperature versus $r_H$ with $\ell = 10$. The solid curve
    represents $T_{H}(r_H,\theta=0.1)$, while the dashed line denotes
    $T_{H}(r_H,\theta\to 0)$ for the NBTZ.}
  \label{fig:T}
\end{figure}

\begin{figure}[pt]
  \includegraphics[width=0.5\textwidth]{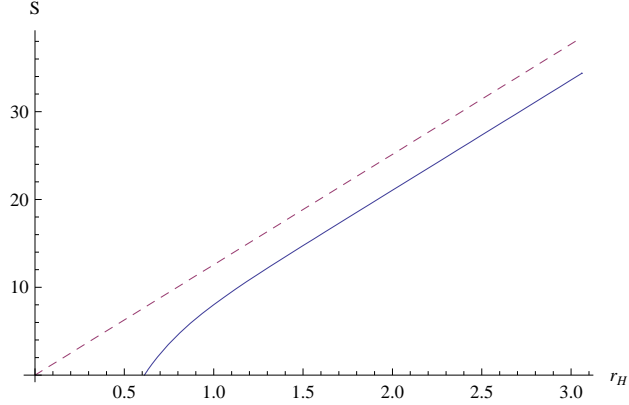}
  \caption{Entropy versus horizon radius $r_H$. The solid curve and dashed line show the entropy $S(r_H,\theta=0.1)$ of
    regular black hole  and $S(r_H,\theta\to 0)=S_{BH}$ of  NBTZ, respectively.}
  \label{fig:S}
\end{figure}

There are two ways to define the entropy. First we introduce the
Bekenstein-Hawking entropy with $G_3=1$ \begin{equation}
S_{BH}=\frac{\pi r_H}{2}.
\end{equation}
Unfortunately, this choice does not satisfy the first-law of
thermodynamics
\begin{equation}
  dM \ne T_H dS_{BH},
\end{equation}
while it satisfies the area-law. On the other hand, we require that
the first-law be satisfied with the regular black hole.  Then, we
obtain the entropy by integrating $dM =T_H dS$ over $r_H$ as
\begin{equation}
  \label{S}
  S(r_H,\theta)  =
  \int_{r_*}^{r_H} \frac{1}{T_H}\Big( \frac{dM}{dr'_H}\Big) dr'_H =
  \frac{\pi^{3/2}}{4} \int_{r_*}^{r_H} \frac{dr'_H}{ \gamma(\frac32,
    \frac{r_H^{'2}}{4\theta})}.
\end{equation}
However, this entropy does not satisfy the area-law. The behavior
of the entropy is depicted in Fig.~\ref{fig:S}. In the limit of
$\theta \to \infty$, one finds that $S(r_H,\theta\to
0)=(\pi/2)\int^{r_H}_0 dr_H'=S_{BH}$.

\begin{figure}[pt]
  \includegraphics[width=0.5\textwidth]{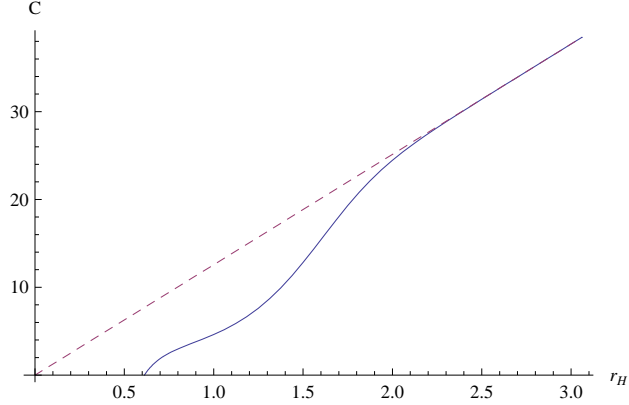}
  \caption{The solid curve and dashed line show the heat capacity $C(r_H,\theta=0.1)$ of
    regular black hole  and $C(r_H,\theta\to 0)$ of  NBTZ, respectively.}
  \label{fig:Ca}
\end{figure}

The heat capacity is defined as
\begin{equation}
  \label{Ca}
  C(r_H,\theta) = \left( \frac{\partial M}{\partial T_H} \right)_{\theta} =
  \left( \frac{\partial M}{\partial r_H} \right)_{\theta} \left(
    \frac{\partial T_H}{\partial r_H} \right)_{\theta}^{-1}.
\end{equation}
The heat capacity determines the thermodynamic stability. For
$C>0$, the black hole is locally stable, while for $C<0$, the
corresponding black hole is locally unstable. As is shown
Fig.~\ref{fig:Ca}, the regular black hole has a single stable
phase, similar to $C(r_H,\theta\to 0)=\pi r_H/2$ of NBTZ. For
large $r_H$, two black holes have the nearly same heat capacity.

\begin{figure}[pt]
  \includegraphics[width=0.5\textwidth]{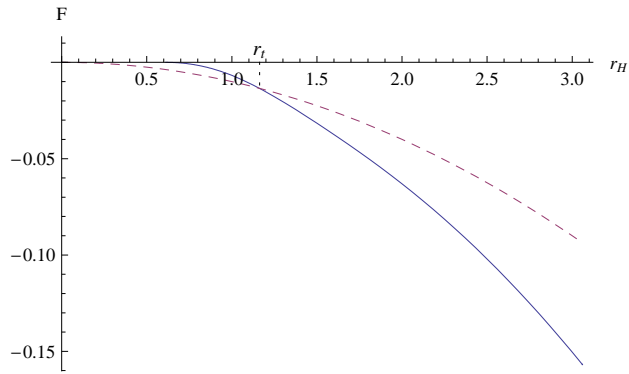}
  \caption{Free Energy versus the horizon radius $r_H$. The solid  and dashed curves show the free energy $F(r_H,\theta=0.1)$ of
    regular black hole  and $F(r_H,\theta\to 0)$ of  NBTZ, respectively. }
  \label{fig:F}
\end{figure}

Finally we define the on-shell free energy
\begin{equation}
  \label{F}
  F(r_H,\theta)= M(r_H,\theta)-M_*(\theta) - T_H(r_H,\theta) S(r_H,\theta).
\end{equation}
Here we use the extremal black hole as the ground
state~\cite{CEJM}, even though there is no  gauge field. This is
mainly because the extremal black hole plays the role of a stable
remnant in the regular black hole. The free energy is shown in
Fig.~\ref{fig:F}. In the limit of $\theta\to 0$, one has the free
energy $F(r_H,\theta\to 0)=-r_H^2/8\ell^2$ of NBTZ. Also we
observe that $F(r_*,\theta)=0$, showing the zero free energy for
the extremal black hole. Importantly, there is a nonvanishing
probability for decay of regular black hole  into NBTZ for
$r_H<r_t$ where $r_t$ is determined by the condition of
$F(r_t,\theta)=F(r_t,\theta\to 0)$, because of
$F(r_t,\theta)>F(r_t,\theta\to 0)$~\cite{MTZ,myung}. On the other
hand, for $r_H>r_t$, there is a nonvanishing probability for decay
of NBTZ to regular black hole because of
$F(r_t,\theta)<F(r_t,\theta\to 0)$.

\section{3D Noncommutative black hole}
\label{sec:gauss} 
\begin{figure}[pt]
  \includegraphics[width=0.5\textwidth]{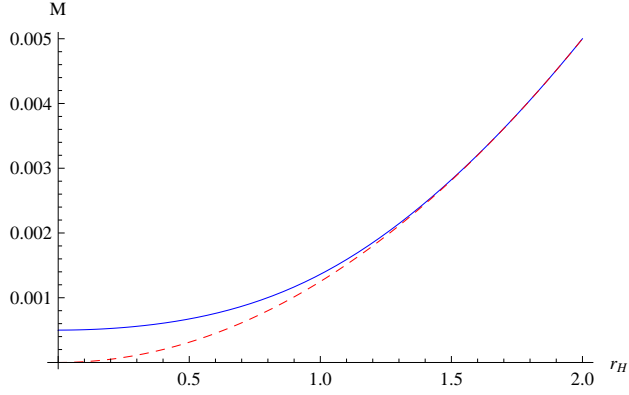}
  \caption{The mass $M$ as a function of horizon radius $r_H$. The
    solid and the dashed line indicate the 3D noncommutative
   black hole  and  NBTZ, respectively. These black holes have single horizon except $r_H=0$.}
  \label{fig:G:M}
\end{figure}

Considering the 3D Gaussian distribution
(\ref{3Dgd}),  the metric function is obtained as
\begin{equation}
  \label{G:metric}
  f^{3D}_\theta(r) = \frac{r^2}{\ell^2} - 8m^{3D}_\theta(r)= \frac{r^2}{\ell^2}-8M\Big(1-e^{-\frac{r^2}{4\theta}}\Big).
\end{equation}
From the condition of $f^{3D}_\theta(r_H)=0$, the mass takes the
form
\begin{equation}
  \label{G:M}
  M = \frac{r_H^2}{8\ell^2 \left[ 1- \exp\left( -\frac{r_H^2}{4\theta}
      \right)\right]}.
\end{equation}
We note that this black hole has single horizon except $r_H=0$.
Observing the metric function (\ref{G:metric}) leads to that
$f^{3D}_\theta \to 0$, as $r \to 0$, irrespective of whatever $M$
is taken. This implies that the mass is not correctly defined in
the limit of $r_H\to 0$. Using the L'Hospital principle, it may
lead to $M \to \frac{\theta}{2\ell^2}$ in the limit of $r_H \to
0$. Thus, we may obtain  Fig.\ \ref{fig:G:M}. However, this value
does not reflect the correct limit because we could not define the
3D noncommutative black hole in the limit of $r_H \to 0$. This
uncertainty propagates all thermodynamic quantities.

\begin{figure}[pt]
  \includegraphics[width=0.5\textwidth]{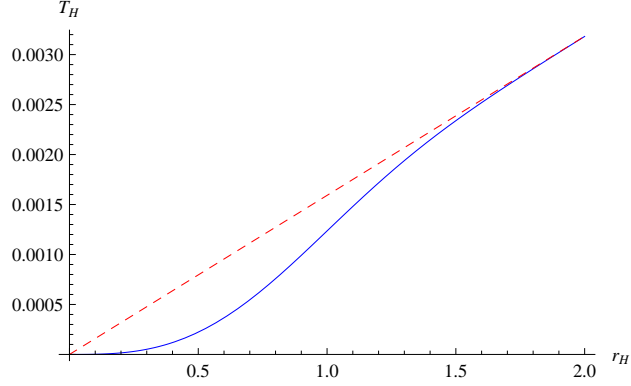}
  \caption{This figure shows the profile of the Hawking
    temperature. The solid and the dashed line indicate the  3D noncommutative
   black hole  and  NBTZ, respectively.}
  \label{fig:G:T}
\end{figure}

The Hawking temperature is computed  to be
\begin{equation}
  \label{G:T}
  T_H = \frac{r_H}{2\pi\ell^2} \left[ 1 + \frac{r_H^2}{4\theta \left(
        1- \exp \left(\frac{r_H^2}{4\theta}\right) \right)}
        \right].
\end{equation}
We note that the temperature takes an indeterminate form of
$\frac{0}{0}$ at $r_H=0$, even though it appears determinate in
Fig.~\ref{fig:G:T}.

\begin{figure}[pt]
  \includegraphics[width=0.5\textwidth]{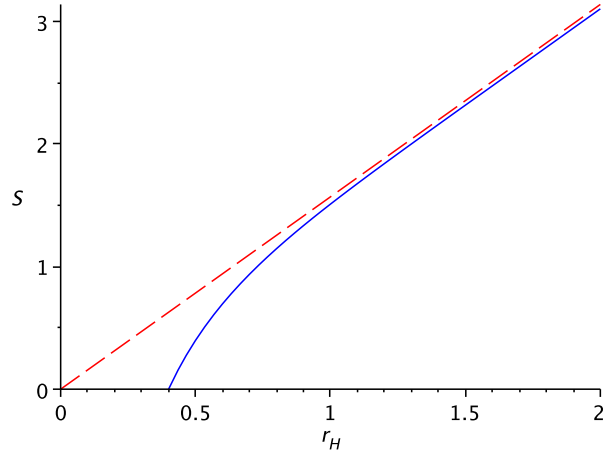}
  \caption{This figure denotes the behavior of the entropy. The solid
    and the dashed line indicate the  3D noncommutative
   black hole  and  NBTZ,
    respectively.}
  \label{fig:G:S}
\end{figure}

The entropy is obtained as
\begin{equation}
  S = \int_{r_0}^{r_H} \frac{dM}{T_H} = \frac{\pi}{2} \int_{r_0}^{r_H}
  \frac{d\xi}{1- \exp \left( - \frac{\xi^2}{4\theta} \right)},
\end{equation}
where $r_0=0.4$ is chosen  for given $\theta=0.1$ and $\ell=10$ by
requiring the condition of  the consistency  with the entropy
$S_{BH}=\pi r_H/2$ of NBTZ  for a large black hole (Fig.\
\ref{fig:G:S}). For a small black hole, the entropy is  not
properly defined as $S \approx 2\pi\theta ( 1/r_0 - 1/r_H)$.

\begin{figure}[pt]
  \includegraphics[width=0.5\textwidth]{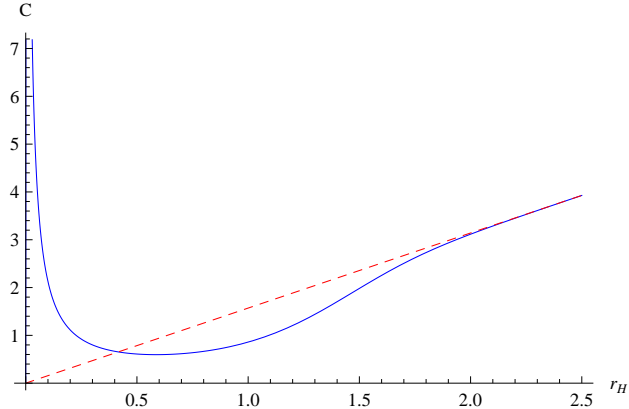}
  \caption{This shows the heat capacity, which is positive
    definite. The solid and the dashed line indicate the  3D noncommutative
   black hole  and  NBTZ, respectively.}
  \label{fig:G:C}
\end{figure}

The heat capacity is calculated as
\begin{eqnarray}
  \label{G:C}
  C &=& \frac{\partial M}{\partial T_H} = \frac{\partial M}{\partial
    r_H} \left(\frac{\partial T_H}{\partial r_H}\right)^{-1} \\
  &=& \frac{\pi r_H}{2} \frac{1-\left( 1+ \frac{r_H^2}{4\theta}
    \right) e^{ - \frac{r_H^2}{4\theta}} }{1 +  e^{ -
      \frac{r_H^2}{4\theta}} \left[ - 2\left(4- \frac{r_H^2}{4\theta}
    \right) \left( 1+ \frac{r_H^2}{4\theta}
    \right) + \left( 1+ \frac{3 r_H^2}{4\theta}
    \right) e^{ - \frac{r_H^2}{4\theta}} \right] },
\end{eqnarray}
which shows an unusual behavior  $C\sim 2\pi\theta/(3r_H)$ for
small $r_H$ (See Fig.~\ref{fig:G:C}), comparing with $C_{NBTZ}=\pi
r_H/2$.

\begin{figure}[pt]
  \includegraphics[width=0.5\textwidth]{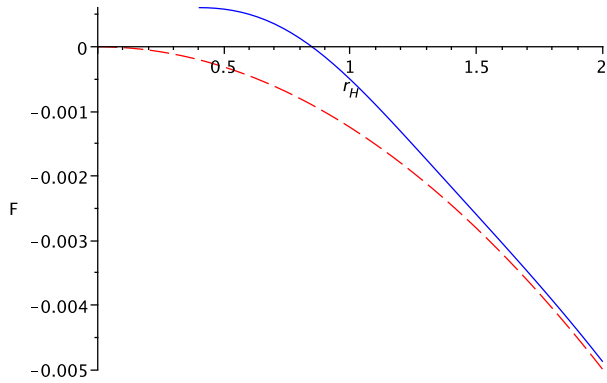}
  \caption{This indicates the free energy. The solid and the dashed line
    indicate the  3D noncommutative
   black hole  and  NBTZ, respectively.}
  \label{fig:G:F}
\end{figure}

The free energy is given by
\begin{equation}
  \label{G:F}
  F = M(r_H) - T_H S(r_H),
\end{equation}
As is shown in Fig.~\ref{fig:G:F}, we could not define the free
energy for small black holes, because there exist uncertainties
for $M$, $T_H$, and $S_{BH}$ for small black holes.

\section{Discussion}
\label{sec:dis}

We construct a regular black hole in AdS$_3$ spacetimes by
introducing an anisotropic perfect fluid (\ref{emt}) inspired by
the 4D noncommutative black hole. This black hole has  a nature of
the 4D noncommutative black hole with two horizons in three
dimensions.

We compare thermodynamics of this black hole with that of
non-rotating BTZ black hole  (NBTZ). The Hawking temperature and
heat capacity of large regular black hole approach those of NBTZ.
However, the entropy of regular black hole is different from the
Bekenstein-Hawking entropy of NBTZ because we use the first-law of
thermodynamics to derive the entropy. Actually, it confirms that
the first-law of thermodynamics is not compatible with the
Bekenstein-Hawking entropy for regular black holes.

From the graph of free energy in Fig.~\ref{fig:F}, we observe that
there is a nonvanishing probability for decay of regular black
hole into NBTZ for $r_H<r_t$, while for $r_H>r_t$, there is a
nonvanishing probability for decay of NBTZ to regular black hole.
This implies that there may exist a phase transition between
regular black hole and NBTZ.

On the other hand, the Gaussian distribution (\ref{3Dgd}) provides
the 3D noncommutative black hole with single horizon except
$r_H=0$. The thermodynamics of small 3D noncommutative black hole
is not well established.  The small 3D noncommutative black hole
is not defined in the limit of $r_H\to 0$ since the a smeared
(Gaussian) distribution around the origin is not appropriate to
make a small black hole, differing from the point Dirac-delta
function.

\section*{Acknowledgments}
This work was supported by the Science Research
Center Program of the Korea Science and Engineering Foundation through
the Center for Quantum Spacetime (CQUeST) of Sogang University with
grant number R11-2005-021.


\end{document}